\documentstyle[12pt]{article}
\newcommand{\be}{\begin{equation}}
\newcommand{\ee}{\end{equation}}
\newcommand{\bea}{\begin{eqnarray}}
\newcommand{\eea}{\end{eqnarray}}
\newcommand{\sptwo}{1.8}
\newcommand{\doublespace}{\edef\baselinestretch{\sptwo}\Large\normalsize}
\begin{document}
\vspace{1.0in}
\begin{center}{\Large\bf On the Interactions of Light Gravitinos}
\end{center}
~\\
\begin{center}T.E. Clark\footnote{e-mail address: clark@physics.purdue.edu}, 
Taekoon Lee\footnote{e-mail address: tlee@physics.purdue.edu}, 
S.T. Love\footnote{e-mail address: love@purdd.physics.purdue.edu}, 
Guo-Hong Wu\footnote{e-mail address: wu@physics.purdue.edu} \\
 ~~ \\
Department of Physics\\
Purdue University\\
West Lafayette, IN 47907-1396
\end{center}
\vspace{1.0in}
\begin{center}
{\bf Abstract}
\end{center}
In models of spontaneously broken supersymmetry, certain light gravitino 
processes are governed by the coupling of its Goldstino components. 
The rules for constructing SUSY and gauge invariant actions 
involving the Goldstino couplings to matter and gauge fields 
are presented. The explicit operator construction is found to be 
at variance with some previously reported claims. A phenomenological 
consequence arising from light gravitino interactions in supernova 
is reexamined and scrutinized.
\pagebreak
\doublespace

In the supergravity theories obtained from gauging a spontaneously 
broken global $N=1$ supersymmetry (SUSY), the Nambu-Goldstone fermion, 
the Goldstino \cite{AV,FI}, 
provides the helicity $\pm \frac{1}{2}$ degrees of 
freedom needed to render the spin 
$\frac{3}{2}$ gravitino massive through the super-Higgs mechanism. 
For a light gravitino, the high energy (well above the gravitino mass) 
interactions of these helicity $\pm \frac{1}{2}$ 
modes with matter will be enhanced according to the supersymmetric 
version of the equivalence theorem \cite{ET}.  The effective action 
describing such interactions can then be constructed using 
the properties of the Goldstino fields. Currently studied gauge mediated 
supersymmetry breaking models $\cite{vis}$ provide a 
realization of this scenario as do certain 
no-scale supergravity models $\cite{nan}$. In the gauge mediated 
case, the SUSY is dynamically broken in a hidden sector of the theory by means 
of gauge interactions resulting in a hidden sector Goldstino 
field. The spontaneous breaking is then mediated to the minimal 
supersymmetric standard model (MSSM) 
via radiative corrections in the standard model gauge interactions 
involving messenger fields which carry standard 
model vector representations.  In such models, 
the supergravity contributions to the SUSY breaking mass splittings are 
small compared to these gauge mediated contributions. Being a gauge singlet, 
the gravitino mass arises only from  
the gravitational interaction and is thus far smaller than the scale 
$\sqrt{F}$, where $F$ is the Goldstino decay constant. Moreover, since the 
gravitino is the lightest of all 
hidden and messenger sector degrees of freedom, 
the spontaneously broken 
SUSY can be accurately described via a non-linear realization. Such a 
non-linear realization of SUSY on the Goldstino fields was originally 
constructed by Volkov and Akulov \cite{AV}.

The leading term in a momentum expansion of the effective action 
describing the Goldstino self-dynamics 
at energy scales below $\sqrt{4\pi F}$ 
is uniquely fixed by the Volkov-Akulov effective 
Lagrangian $\cite{AV}$ which takes the 
form
\be \label{LAV}
{\cal{L}}_{AV}=-\frac{F^2}{2}~\det{A}.
\ee
Here the Volkov-Akulov vierbein is defined as 
$A_\mu~^\nu = \delta_\mu^\nu + \frac{i}{F^2}\lambda 
\stackrel{\leftrightarrow}{\partial}_\mu\sigma^\nu\bar{\lambda}$, 
with $\lambda (\bar{\lambda})$ the Goldstino Weyl spinor field. 
This effective Lagrangian provides a 
valid description of the Goldstino self interactions 
independent of the particular (non-perturbative) 
mechanism by which the SUSY is dynamically broken. 
The supersymmetry transformations are 
nonlinearly realized on the Goldstino fields as 
$\delta^Q(\xi,\bar{\xi})\lambda^\alpha= F\xi^\alpha +\Lambda^\rho \partial_\rho 
\lambda^\alpha~~;~~\delta^Q(\xi,\bar{\xi})\bar{\lambda}_{\dot\alpha}= 
F\bar\xi_{\dot\alpha} 
+\Lambda^\rho \partial_\rho \bar\lambda_{\dot\alpha}$, 
where $\xi^\alpha,\bar{\xi}_{\dot\alpha}$ are Weyl spinor 
SUSY transformation parameters and $\Lambda^\rho \equiv 
-\frac{i}{F}\left( \lambda\sigma^\rho \bar\xi 
-\xi \sigma^\rho \bar\lambda \right)$ is a 
Goldstino field dependent translation vector. 
Since the Volkov-Akulov Lagrangian transforms as the total divergence 
$\delta^Q(\xi,\bar{\xi}){\cal{L}}_{AV} = \partial_\rho \left( \Lambda^
\rho 
{\cal L}_{AV} \right)$, 
the associated action $I_{AV}=\int d^4x ~{\cal{L}}_{AV}$
is SUSY invariant.

The supersymmetry algebra can also be nonlinearly realized on 
the matter (non-Goldstino) fields, 
generically denoted by $\phi^i$, where $i$ can represent 
any Lorentz or internal symmetry labels, as 
\be
\delta^Q(\xi,\bar{\xi})\phi^i=\Lambda^\rho \partial_\rho \phi^i ~~.
\ee 
This is referred to as the standard realization 
\cite{W}-\cite{CL2}. It can be used, 
along with space-time translations, to readily establish the SUSY 
algebra. Under the non-linear SUSY standard realization, 
the derivative of a matter field transforms as 
$\delta^Q(\xi,\bar{\xi})(\partial_\nu\phi^i)=
\Lambda^\rho \partial_\rho ( \partial_\nu \phi^i ) 
+(\partial_\nu \Lambda^\rho ) 
( \partial_\rho \phi^i )$. 
In order to eliminate the second term on the right hand side 
and thus restore the standard SUSY realization, 
a SUSY covariant derivative is introduced and 
defined so as to transform analogously to $\phi^i$. 
To achieve this, we use the transformation property 
of the Volkov-Akulov vierbein and define the non-linearly realized 
SUSY covariant derivative \cite{CL2}
\be
\label{SCD}
{\cal{D}}_\mu\phi^i=(A^{-1})_\mu~^\nu \partial_\nu\phi^i~~,
\ee
which varies according to the standard realization of SUSY: \\
$\delta^Q(\xi,\bar{\xi})({\cal{D}}_\mu\phi^i) =\Lambda^\rho 
\partial_\rho \left( {\cal{D}}_\mu \phi^i 
\right)$. 

Any realization of the SUSY transformations can be converted to the 
standard realization.  In particular, consider the 
gauge covariant derivative, 
\be
\label{coderiv}
(D_\mu \phi)^i \equiv \partial_\mu\phi^i +{T}^a_{ij}{A}^a_\mu \phi^j~~,
\ee
with $a=1,2,\ldots , {\rm Dim}~{\cal G}$. We seek a SUSY and gauge 
covariant derivative $({\cal D}_\mu \phi)^i$, which transforms as the 
SUSY standard realization. Using the Volkov-Akulov vierbein, we define 
\be 
({\cal D}_\mu \phi)^i \equiv (A^{-1})_\mu~^\nu (D_\nu \phi)^i~~,
\ee
which has the desired transformation property, 
$\delta^Q(\xi,\bar{\xi})({\cal D}_\mu \phi)^i =\Lambda^\rho\partial_
\rho ({\cal D}_\mu \phi)^i$, provided 
the vector potential has the SUSY transformation 
$\delta^Q(\xi,\bar{\xi})A_\mu \equiv 
\Lambda^\rho \partial_\rho  A_\mu +\partial_\mu \Lambda^\rho 
A_\rho$. Alternatively, we can introduce a redefined gauge field
\be
V_\mu^a \equiv (A^{-1})_\mu~^\nu A_\nu^a~~,
\ee
which itself transforms as the standard realization, 
$\delta^Q(\xi,\bar{\xi})V_\mu^a  = \Lambda^\rho\partial_\rho V_\mu^a$,  and 
in terms of which the standard realization SUSY and gauge 
covariant derivative then takes the form 
\be
({\cal D}_\mu \phi)^i \equiv (A^{-1})_\mu~^\nu 
\partial_\nu \phi^i + T^a_{ij} V^a_\mu 
\phi^j ~~. 
\ee

Under gauge transformations parameterized by $\omega^a$, the 
original gauge field varies as 
$\delta^G (\omega) A_\mu^a  = (D_\mu \omega)^a 
 =\partial_\mu \omega^a +gf_{abc}A_\mu^b \omega^c$, 
while the redefined gauge field $V_\mu^a$ has the Goldstino dependent 
transformation:  
$\delta^G(\omega) V_\mu^a  = (A^{-1})_\mu~^\nu (D_\nu \omega)^a$. 
For all realizations, the gauge transformation and 
SUSY transformation commutator yields a gauge 
variation with a SUSY transformed value of the gauge transformation 
parameter,
\be
\left[\delta^G(\omega),\delta^Q(\xi, \bar\xi)\right]=
\delta^G(\Lambda^\rho \partial_\rho \omega-\delta^Q(\xi, \bar\xi)\omega)~~.
\ee
If we further require the local gauge transformation parameter to 
also transform under the standard 
realization so that 
$\delta^Q(\xi,\bar{\xi}) \omega^a  = \Lambda^\rho\partial_\rho \omega^a$, then 
the gauge and SUSY transformations commute.

In order to construct an invariant kinetic energy term for the 
gauge fields, it is convenient for the gauge covariant 
anti-symmetric tensor field strength to also be brought into the 
standard realization. The usual field strength $F_{\alpha\beta}^a = 
\partial_\alpha A_\beta^a -\partial_\beta A_\alpha^a +if_{abc}A_\alpha^b 
A_\beta^c$ varies under SUSY transformations as 
$\delta^Q(\xi,\bar{\xi}) F_{\mu\nu}^a = 
\Lambda^\rho\partial_\rho F_{\mu\nu}^a  +\partial_\mu 
\Lambda^\rho F_{\rho\nu}^a +\partial_\nu \Lambda^\rho F_{\mu\rho}^a$. 
A standard realization of the gauge covariant 
field strength tensor, ${\cal F}_{\mu\nu}^a$, can be then defined as 
\be
\label{field}
{\cal F}_{\mu\nu}^a = 
(A^{-1})_\mu~^\alpha (A^{-1})_\nu~^\beta F_{\alpha\beta}^a~~,
\ee
so that $\delta^Q(\xi,\bar{\xi}) {\cal F}_{\mu\nu}^a = 
\Lambda^\rho\partial_\rho {\cal F}_{\mu\nu}^a$. 

These standard realization building blocks consisting of 
the gauge singlet Goldstino SUSY covariant derivatives, 
${\cal D}_\mu \lambda,~ {\cal D}_\mu 
\bar\lambda$, the matter fields, 
$\phi_i$, their SUSY-gauge covariant derivatives, 
${\cal D}_\mu \phi^i$, and the field strength tensor, 
${\cal F}_{\mu\nu}^a$, along with their higher covariant derivatives 
can be  combined to make SUSY and gauge invariant actions. These invariant 
action terms then dictate the couplings of the Goldstino which, in general,  
carries the residual consequences of the spontaneously broken supersymmetry.

A generic SUSY and gauge invariant action can be constructed $\cite{CL2}$ as 
\be
\label{IEFF}
I_{\rm eff}=\int d^4x \,\, det A \,\, {\cal L}_{\rm eff}({\cal D}_\mu \lambda, 
{\cal D}_\mu \bar{\lambda}, \phi^i, {\cal D}_\mu \phi^i, {\cal F}_{\mu\nu})
\ee
where ${\cal L_{\rm eff}}$ is any gauge invariant 
function of the standard realization basic building blocks. Using the 
nonlinear SUSY transformations 
$\delta^Q(\xi,\bar{\xi}) \, det A = \partial_\rho (\Lambda^\rho \, det A)$ and 
$\delta^Q(\xi,\bar{\xi}) {\cal L}_{\rm eff}= \Lambda^\rho \partial_\rho 
{\cal L}_{\rm eff}$, 
it follows that $\delta^Q(\xi,\bar{\xi}) I_{\rm eff}=0$. 

It proves convenient to catalog the terms in the effective Lagranian, 
${\cal L}_{\rm eff}$, by an expansion in the number of 
Goldstino fields which appear when covariant derivatives are replaced by 
ordinary derivatives and the Volkov-Akulov vierbein appearing in the standard 
realization field strengths are set to unity.  
So doing, we expand
\be
{\cal L}_{\rm eff}= \left[{\cal L}_{(0)}+ {\cal L}_{(1)}+ 
{\cal L}_{(2)}+\cdots \right] ~,
\ee
where the subscript $n$ on ${\cal L}_{(n)}$ denotes that each independent 
SUSY invariant operator in 
that set begins with $n$ Goldstino fields.  

${\cal L}_{(0)}$  consists of all gauge and SUSY invariant operators 
made only from light matter 
fields and their SUSY covariant derivatives.  Any Goldstino field 
appearing in ${\cal L}_{(0)}$ arises only from 
higher dimension terms in the matter covariant derivatives and/or the field 
strength tensor. Taking the light non-Goldstino fields to be those of the 
MSSM and retaining terms through 
mass dimension 4, then ${\cal L}_{(0)}$ is well 
approximated by the Lagrangian of the minimal supersymmetric 
standard model which includes the soft SUSY breaking terms, 
but in which all derivatives have been replaced by SUSY 
covariant ones and the field strength tensor replaced by the 
standard realization field strength: 
\be
{\cal L}_{(0)} = {\cal L}_{MSSM}(\phi,{\cal D}_\mu \phi, {\cal F}_
{\mu\nu}) .
\ee
Note that the 
coefficients of these terms are fixed by the normalization of the 
gauge and matter fields, their masses and self-
couplings; that is, the normalization of the Goldstino independent Lagrangian.

The ${\cal L}_{(1)}$ terms in the effective Lagrangian begin with 
direct coupling of one Goldstino covariant derivative to the 
non-Goldstino fields.  The general form of these terms, retaining operators 
through mass dimension 6, is given by
\be
{\cal L}_{(1)}= \frac{1}{F}[{\cal D}_\mu \lambda^\alpha Q_{MSSM \alpha}^\mu 
+ \bar 
Q_{MSSM \dot\alpha}^\mu {\cal D}_\mu \bar\lambda^{\dot\alpha}] ,
\ee
where $ Q_{MSSM \alpha}^\mu$ and $\bar Q_{MSSM \dot\alpha}^\mu$ 
contain the pure MSSM field contributions to 
the conserved gauge invariant supersymmetry currents with once again 
all field derivatives being replaced by SUSY covariant 
derivatives and the vector field strengths in the standard realization. 
That is, it is this term in the effective Lagrangian which, using 
the Noether construction, produces the Goldstino independent piece 
of the conserved supersymmetry current. 
The Lagrangian ${\cal L}_{(1)}$ describes processes involving 
the emission or absorption of a single helicity $\pm \frac{1}{2}$ gravitino. 

Finally the remaining terms in the effective Lagrangian all contain  
two or more Goldstino fields.  In 
particular, ${\cal L}_{(2)}$ begins with the coupling of two Goldstino 
fields to matter or gauge fields. Retaining terms through mass dimension 8 
and focusing only on the $\lambda-\bar{\lambda}$ terms, we can write   
\bea
{\cal L}_{(2)} &=& \frac{1}{F^2}{\cal D}_\mu \lambda^\alpha {\cal 
D}_\nu 
\bar\lambda^{\dot\alpha} M^{\mu\nu}_{1\alpha\dot\alpha} +\frac{1}{
F^2}
{\cal D}_\mu \lambda^\alpha \stackrel{\leftrightarrow}{\cal D}_\rho{
\cal D}_\nu 
\bar\lambda^{\dot\alpha} M^{\mu\nu\rho}_{2\alpha\dot\alpha}\cr
 & &  + \frac{1}{F^2}{\cal D}_\rho
\left[ {\cal D}_\mu \lambda^\alpha {\cal D}_\nu 
\bar\lambda^{\dot\alpha}\right] M^{\mu\nu\rho}_{3\alpha\dot\alpha},
\eea
where the standard realization composite operators that contain matter 
and gauge fields are denoted by the 
$M_i$.  They can be enumerated by their operator dimension, Lorentz 
structure and field content. In the gauge mediated 
models, these terms are all generated by 
radiative corrections involving the standard model gauge coupling 
constants. 

Let us now focus on the pieces of ${\cal L}_{(2)}$ which contribute to 
a local operator containing two gravitino fields and 
is bilinear in a Standard Model fermion $(f, \bar{f})$. 
Those lowest dimension operators (which  
involve no derivatives on $f$ or $\bar{f}$) 
are all contained in the $M_1$ piece.  
After application of the Goldstino field equation (neglecting the 
gravitino mass) and making prodigious use of Fierz rearrangement identities,  
this set reduces to just 1 independent on-shell interaction term. 
In addition to this operator, there 
is also an operator bilinear in $f$ and $\bar{f}$ 
and containing 2 gravitinos which arises from 
the product of $det~ A$ with ${\cal L}_{(0)}$. Combining the two 
independent on-shell interaction terms involving 2 gravitinos and 2 fermions, 
results in the effective action 
\bea
I_{f\bar f \tilde{G}\tilde{G}} &=& \int d^4x \left[-\frac{1}{2F^2}
\left( \lambda \stackrel{\leftrightarrow}{\partial}_\mu 
\sigma^\nu \bar\lambda\right) \left( f 
\stackrel{\leftrightarrow}{\partial}_\nu \sigma^\mu \bar f \right) 
\right.\cr
 &+& \left. \frac{C_{ff}}{F^2}\left( f\partial^\mu \lambda\right)\left(\bar 
f \partial_\mu \bar\lambda\right) \right] 
~~,
\eea
where $C_{ff}$ is a model dependent real coefficient. Note that 
the coefficient of the 
first operator is fixed by the normaliztion of the MSSM Lagrangian. This 
result is in accord with a recent analysis \cite{Zw} where it was found that 
the fermion-Goldstino scattering amplitudes depend on only one parameter 
which corresponds to the coefficient $C_{ff}$ in our notation.

In a similar manner, the lowest mass dimension operator contributing to the  
effective action describing the coupling of two on-shell gravitinos 
to a single photon arises from the $M_1$ and $M_3$ pieces of ${\cal L}_{(2)}$ 
and has the form
\be
\label{gamma}
I_{\gamma\tilde{G}\tilde{G}}=\int d^4x \left[\frac{C_\gamma}{F^2} \left( 
\partial^\mu \lambda \sigma^\rho \partial^\nu \bar\lambda \right) 
\partial_\mu F_{\rho\nu}\right] + h.c.~~,
\ee
with $C_\gamma$ a model dependent real coefficient and $F_{\mu\nu}$ is the 
electromagnetic field strength. Note that the operator in the 
square bracket is odd under both parity ($P$) and charge conjugation ($C$). 
In fact any operator arising from a gauge and SUSY invariant structure 
which is bilinear in two on-shell gravitinos and contains only a 
single photon is necessarily odd in both $P$ and $C$. 
Thus the generation of any such operator 
requires a violation of both $P$ and $C$. 
Using the Goldstino equation of motion, 
the analogous term containing $\tilde{F}_{\mu\nu}$  
reduces to Eq.(\ref{gamma}) with $C_\gamma \rightarrow -iC_\gamma$. 
Recently, there has appeared in the 
literature \cite{Luty} the claim that there is a lower dimensional operator 
of the form $\frac{\tilde{M}^2}{F^2}
\partial^\nu\lambda \sigma^\mu \bar{\lambda}  F_{\mu\nu}$ which 
contributes to the single photon- 2 gravitino interaction. Here 
${\tilde{M}}$ is a model dependent SUSY breaking mass parameter 
which is roughly an order(s) of magnitude less than $\sqrt{F}$. 
>From our analysis, we do not find such a term to be part of a 
SUSY invariant action piece and thus it should not be included in the 
effective action. Such a term is also absent if one employs 
the equivalent formalism of Wess and Samuel \cite{W}. 
We have also checked that such a term does not 
appear via radiative corrections by 
an explicit graphical calculation using the correct non-linearly 
realized SUSY invariant action. This is 
also contrary to the previous claim. 

There have been several recent attempts to extract a lower bound on 
the SUSY breaking scale using the supernova cooling rate 
\cite{Luty,moha,moha1}. 
Unfortunately, some of these estimates \cite{Luty,moha1} 
rely on the existence of the 
non-SUSY invariant dimension 6 operator referred to above. Using the 
correct low energy effective lagrangian 
of gravitino interactions, the leading term coupling 2 gravitinos to a 
single photon contains an additional supression factor of roughly 
$C_\gamma\frac{s}{\tilde{M}^2}$. 
Taking $\sqrt{s}\simeq 0.1~{\rm GeV}$ for the processes of 
interest and using $\tilde{M}\sim 100~{\rm GeV}$, this introduces an 
additional supression of at least $10^{-12}$ in the rate 
and obviates the previous estimates of a bound on $F$. 

Assuming that the mass scales of gauginos and the superpartners of 
light fermions are above the core temperature of supernova, 
the gravitino cooling of supernova
occurs mainly via  gravitino pair production. It is interesting to 
compare the gravitino pair production cross section to that of the neutrino 
pair production, which is the main supernova cooling channel. 
We have seen that for low energy gravitino interactions with 
matter, the amplitudes for gravitino 
pair production is proportional to $ 1/F^{2}$. 
A simple dimensional analysis then suggests the ratio of the cross 
sections is:
\be
\frac{\sigma_{\chi\chi}}{\sigma_{\nu\nu}} \sim \frac{ s^{2}}{F^{4}
G_{\mbox{\tiny F}}^{2}}
\ee
where $G_{\mbox{\tiny F}}$ is the Fermi coupling and 
$\sqrt{s}$ is the typical energy scale of the particles in a supernova.
Even with the most optimistic values for $F$, the gravitino production 
is too small to be relevant. For example, taking $\sqrt{F}=100\,\, GeV, \sqrt{
s}=.1\,\, GeV$,
the ratio is of $O(10^{-11})$. It seems, therefore, that such 
an astrophysical bound on the SUSY breaking scale is untenable in models where 
the gravitino is the only superparticle below the scale of supernova 
core temperature. 
\\

\noindent
We thank T. K. Kuo for useful conversations. 
This work was supported in part by the U.S. Department of Energy 
under grant DE-FG02-91ER40681 
(Task B).
\pagebreak

\newpage
\end{document}